\documentclass[conference]{IEEEtran}
\IEEEoverridecommandlockouts
\usepackage{cite}
\usepackage{amsmath,amssymb,amsfonts}
\usepackage{graphicx}
\usepackage{textcomp}
\def\BibTeX{{\rm B\kern-.05em{\sc i\kern-.025em b}\kern-.08em
    T\kern-.1667em\lower.7ex\hbox{E}\kern-.125emX}}

\usepackage[usenames,dvipsnames,svgnames,xcdraw,table]{xcolor}
\usepackage[utf8x]{inputenc}
\usepackage{listings}

\usepackage{caption}
\usepackage{subcaption}
\usepackage{paralist}
\usepackage{multirow}

\usepackage[linesnumbered,ruled,vlined]{algorithm2e}
\SetKwInput{KwInput}{Input}                
\SetKwInput{KwOutput}{Output}              

\definecolor{codegreen}{rgb}{0,0.6,0}
\definecolor{codegray}{rgb}{0.5,0.5,0.5}
\definecolor{codepurple}{rgb}{0.58,0,0.82}
\definecolor{backcolour}{rgb}{0.95,0.95,0.92}
\definecolor{darkblue}{rgb}{0.0,0.0,0.6}
\definecolor{modebeige}{rgb}{0.59, 0.44, 0.09}
\definecolor{sepia}{rgb}{0.44, 0.26, 0.08}
\definecolor{olivedrab7}{rgb}{0.24, 0.2, 0.12}

\newif\ifComments
\Commentstrue
\newcommand{\REM}[1]{}

\ifComments
\newcommand{\juan}[1]{\noindent\textcolor{green}{Juan: {#1}}}
\newcommand{\alex}[1]{\noindent\textcolor{magenta}{Alexandro: {#1}}}
\newcommand{\rem}[1]{\noindent\textcolor{brown}{Removed: {#1}}}

\newcommand{\ed}[1]{\noindent\textcolor{red}{ {#1}}}
\else
\newcommand{\juan}[1]{}
\newcommand{\alex}[1]{}
\newcommand{\rem}[1]{}

\newcommand{\ed}[1]{}
\fi

\begin{document}

\title{Evaluating the Performance of Speculative DOACROSS Loop Parallelization with {\tt taskloop}
\thanks{This work is supported by FAPESP (grants 18/07446-8 and 18/15519-5).}
}

\author{\IEEEauthorblockN{Juan Salamanca}
\IEEEauthorblockA{\textit{DEMAC -- São Paulo State University}\\
juan@rc.unesp.br}
\and
\IEEEauthorblockN{Alexandro Baldassin}
\IEEEauthorblockA{\textit{DEMAC -- São Paulo State University}\\
alex@rc.unesp.br}
}

\maketitle

\begin{abstract}
OpenMP provides programmers with directives to parallelize DOALL loops such as {\tt parallel for} and, more recently, {\tt taskloop} for task-based parallelism. On the other hand, when it is possible to prove that a loop is DOACROSS, programmers can try to parallelize it through {\tt parallel for} and to use the OpenMP {\tt ordered} directive to mark the region of the loop that has to be executed sequentially. However, when neither of the previous two cases can be proven, programmers have to be conservative and assume that the loop is DOACROSS (actually {\it may} DOACROSS). Previous work proposed speculative support for {\tt taskloop} ({\tt tls} clause) and thus made it possible to parallelize {\it may} DOACROSS loops exploiting task-based parallelism and the fact that many of them are computationally intensive and DOALL at runtime. This paper proposes Speculative Task Execution (STE) through the addition of speculative privatizations to {\tt taskloop tls} with two novel clauses: {\tt spec\_private} and {\tt spec\_reduction}. We also present a performance comparison between {\tt taskloop-tls} with speculative privatizations vs. {\tt ordered} that reveals that, for certain loops, slowdowns using OpenMP DOACROSS can be transformed in speed-ups of up to $1.87\times$ by applying speculative parallelization of tasks.
\end{abstract}

\begin{IEEEkeywords}
Thread-Level Speculation, OpenMP taskloop, DOACROSS
\end{IEEEkeywords}

\section{Introduction}\label{sec:introduction}


Loops account for most of the execution of programs and much research has been dedicated to parallelize the iterations of loops~\cite{Lamport1974,CytronICPP86,\REM{AikenESOP88,}HursonAC97,\REM{OttoniMICRO05,}VachharajaniPACT07}. However, the biggest challenge for the compiler is to statically analyze loops and determine if they do not have loop-carried dependences (DOALL). Often, this task is prevented by the use of pointers and the indeterminism of the control flow.

Many performant loop parallelization techniques are typically restricted to DOALL loops or loops that contain well-known dependence patterns (e.g., reduction). This restriction precludes the parallelization of many computational intensive non-DOALL loops. In such loops, either the compiler finds at least one loop-carried dependence (DOACROSS loop) or it cannot prove, at compile-time, that the loop is free of such dependences, even though they might never show up at runtime ({\it may} DOACROSS loop). In any case, most compilers end up not parallelizing non-DOALL loops.

In addition to loops, there are other hot-code regions that the programmer or the compiler tries to parallelize. The task-based programming model provides annotations of dependences between these code regions (tasks). Thus, a runtime system manages these dependences and schedules tasks to execute on cores.  \REM{Differently from thread parallelism, task parallelism does not focus on mapping parallelism to threads, but it is oblivious of the physical layout and focuses on exposing more parallelism. Task parallelism was implemented to be more versatile than thread-level parallelism~\cite{AyguadeTPDS09} and was added to OpenMP in version 3.0.} OpenMP 4.5~\cite{openmp2015} added several constructs to users as explained in Section~\ref{sec:background}. One of these constructs is {\tt taskloop}, which allows programmers to parallelize loops in a similar fashion to the {\tt parallel for} construct. It creates explicit tasks rather than implicit ones (as in {\tt omp for}) and divides the loop iteration space between them~\cite{TeruelIWOMP13}. However, the loop to be parallelized using {\tt taskloop} has to be DOALL or contain well-known reduction patterns\REM{(the {\tt reduction} clause in tasks was recently added to OpenMP 5.0~\cite{openmp2018})}.

A technique to enable parallel execution in the presence of potential dependences is \emph{Thread-Level Speculation} (TLS)~\cite{SohiISCA95,SteffanHPCA98}\REM{,SteffanISCA2000,SteffanTCS2005}. We proposed the use of Hardware-Transactional-Memory-based TLS (TLS-HTM) to tasks --- {\em Speculative Task Execution} (STE) --- through the addition of the clause {\tt tls} to {\tt taskloop} in a previous work~\cite{SalamancaIWOMP19}. This clause can be used to speculate data dependences between tasks  generated by a {\tt taskloop} construct in non-DOALL loops, thus STE manipulates multiple tasks of loop iterations in order to exploit speculative task parallelism and to accelerate the loop execution.  In this paper, we propose the addition of speculative privatizations to {\tt taskloop tls} through two novel clauses --- {\tt spec\_private} and {\tt spec\_reduction} --- to integrate speculative execution into OpenMP task-based parallelization.

\begin{figure*}[t]
\vspace{-3mm}
\centering
\begin{minipage}{.47\textwidth}
\centering
\begin{lstlisting}[escapeinside={\%*}{*)}] 
   #pragma omp parallel num_threads(N_CORES)
   #pragma omp for ordered(1) shared(n)...
   for(i=5;i<y_size-5;i++){//loopV
      #pragma omp ordered depend (sink:i-1)
      for(j=5;j<x_size-5;j++){
         x = r[i][j];
         if (x>0 &&(/*compare x*/)){ 
            corner_list[n].info=0; %*\label{line:read_n}*)
            corner_list[n].x=j; 
            ...
            n++;       
         }
      }
      #pragma omp ordered depend (source)
   }  
\end{lstlisting}
\vspace{-3mm}
\caption{Fragment of {\tt susan\_corners}'s loop ({\tt loopV}) using {\tt ordered depend}}
\label{lst:loopV}
\end{minipage}
\hfill
\begin{minipage}{.47\textwidth}
\centering
\begin{lstlisting}[escapeinside={\%*}{*)}] 
   #pragma omp parallel num_threads(N_CORES)
   #pragma omp single
   #pragma omp taskloop tls(STRIP_SIZE) spec_private(n)...
   for(i=5;i<y_size-5;i++){//loopV
      for(j=5;j<x_size-5;j++){
         x = r[i][j];
         if (x>0 &&(/*compare x*/)){
            #pragma tls if_read(n) 
            corner_list[n].info=0; %*\label{line:read_n2}*)
            corner_list[n].x=j;        
            ...
            n++;
            #pragma tls if_write(n) 
         }
      } 
   }
\end{lstlisting}
\vspace{-3mm}
\caption{The same loop using {\tt tls} clause and {\tt taskloop}}
\label{lst:loopV_tls}
\end{minipage}
\hfill
\vspace{-4mm}
\end{figure*}

To explain such approach, Figure~\ref{lst:loopV} lists the code of a {\tt for} loop ({\tt loopV}) from {\tt susan\_corners} benchmark where, depending on the value of variable  {\tt x}, it updates a position of an array of corners indexed by {\tt n} and increases by one the value of the variable {\tt n}. There is a {\it may} loop-carried dependence in Line~\ref{line:read_n} because this statement could use a value of {\tt n} calculated in earlier iterations depending on the {\tt if} condition; thus, the use of {\tt taskloop} could violate some loop-carried dependence. Moreover, according to OpenMP specification, this loop is non-conforming for {\tt taskloop} because it relies on the execution order of the iterations and must be serialized. Thus a compiler or programmer gives up parallelizing this loop, or must use DOACROSS (e.g., OpenMP {\tt ordered} construct as shown in Fig.~\ref{lst:loopV}), yielding slowdowns with respect to the serial execution. However, this loop can be speculatively parallelized using STE ({\tt taskloop tls}) as shown in Fig.~\ref{lst:loopV_tls}. Using STE, tasks are generated by grouping loop iterations and then their data dependences are speculated relying on the TLS-HTM mechanisms to detect conflicts and to rollback speculative transactions, as explained in~\cite{SalamancaTPDS18}.

\begin{figure}[b]
\vspace{-4mm}
\centering
\includegraphics[scale=0.5]{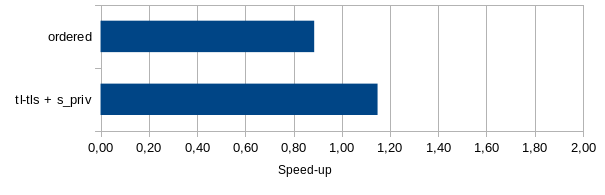}
\vspace{-3mm}
\caption{Performance of {\tt loopV} using: (a) {\tt ordered}; and (b) {\tt taskloop tls} with speculative privatizations}
\label{fig:intro}
\vspace{-1mm}
\end{figure}

Fig.~\ref{fig:intro} shows the speed-ups (with respect to sequential execution and using four threads) of {\tt loopV} (compiled with Clang and linked against the Intel OpenMP runtime) for the following cases: (a) when using the {\tt ordered} clause (above); and (b) when using {\tt taskloop} and the proposed {\tt tls} and {\tt spec\_private} clauses (below).  Speed-ups measurements were performed in a quad-core Intel Skylake machine with TSX-NI support. As shown, {\tt ordered} serializes the execution of the iterations resulting in performance degradation due to the synchronization overhead. On the other hand, {\tt taskloop tls} parallelization results in an improvement of 15\% with respect to serial.

In this paper we make the following contributions:

\begin{inparaitem}
\noindent\item We propose Speculative Task Execution through {\tt tls} and two novel clauses: {\tt spec\_private} and {\tt spec\_reduction}. Together, they extend the {\tt taskloop} construct and enables the programmer to parallelize ({\it may}) DOACROSS loops using TLS-HTM (Section~\ref{sec:proposal});

\noindent\item We perform an evaluation of the proposed clauses. The experimental results show their effectiveness. We further compare against the {\tt ordered} construct, which was implemented previously in~\cite{ShirakoIWOMP13} (Section~\ref{sec:experiments}).
\end{inparaitem}

This paper is organized as follows. Section~\ref{sec:background} describes the background material and discusses related works. Section~\ref{sec:proposal} details the design  and implementation of STE through the new clauses. Benchmarks, methodology and settings are described in Section~\ref{sec:setup}. Section~\ref{sec:experiments}  evaluates the performance of STE. 
Finally, Section~\ref{sec:conclusions} concludes the work.

\section{Background and Related Work}\label{sec:background}

This section presents related works and the main concepts used in this paper: DOACROSS parallelization in OpenMP, Task-based Parallelism, OpenMP tasks, Thread-Level Speculation on Hardware Transactional Memory, and Speculative Privatization.

\subsection{DOACROSS in OpenMP}\label{sec:ordered}

DOACROSS (proposed by Cytron et al.~\cite{CytronICPP86}) is a parallelizing algorithm for loops with loop-carried dependences. The main idea is to distribute the iterations cyclically among the threads trying to simultaneously execute many iterations. As the algorithm makes an effort to parallelize loops with loop-carried dependences, inter-thread communication is required to forward dependences and synchronize shared resources.


DOACROSS support in OpenMP is a variation of the DOACROSS algorithm~\cite{CytronICPP86} and was proposed by Shirako {\it  et al.}~\cite{ShirakoIWOMP13} as an OpenMP construct, which was implemented in OpenMP 4.5~\cite{openmp2015}.  This construct, {\tt ordered}, is used to annotate sequential loop components so as to enable fine-grained parallelism. 

OpenMP 4.0 introduced the {\tt ordered} construct for use in loops marked with {\tt ordered} clause and thus enabling the serial execution of the region within this construct (following the order of loop iterations). However, in OpenMP 4.5, it is possible to use {\tt ordered} as a stand-alone directive that specifies loop-carried dependences. This directive sequentializes and orders the execution of the regions marked among {\tt ordered} while the code outside can run in parallel. If any {\tt depend} clause is specified (as in Figure~\ref{lst:loopV}) then it gives the information of the order in which the threads execute {\tt ordered} regions. In the example, the first {\tt ordered} directive tells the runtime to wait for the culmination of the specified earlier iterations, and the last {\tt ordered} directive marks the point where the iteration computed data that other iterations might need.

\subsection{Task-based Parallelism}

In this model, the execution can be modeled as a directed acyclic graph, where nodes are tasks and edges define data dependences between tasks.\REM{A runtime system schedules tasks whose dependences are resolved over available worker threads thus enabling load balancing and work stealing\REM{~\cite{GayatriHIPC13}}.}
At runtime, the task creation code packs the kernel code pointer and the task operands, and then adds them in the task pipeline; in this way, the generating thread can continue creating additional tasks. The pipeline decodes task dependences, generates the dependence graph using this information, and schedules tasks when they are ready over available worker threads thus enabling load balancing and work stealing~\cite{GayatriHIPC13\REM{,PerezCLUSTER08}}. \REM{StarSs has implementations for multicore architectures such as Symmetric Multiprocessors (SMP), the Cell Broadband Engine, Graphical Processing Units (GPU) and clusters. SMP Superscalar (SMPSs) is an implementation of StarSs for SMP.}

\REM{To explore task-based programming models, OpenMP and Intel TBB are increasing their popularity thus confirming that the task abstraction is an intuitive construct.}
\REM{StarSs is a programming model  which  supports out-of-order execution of tasks by enabling the programmer to identify data dependences between tasks through annotations of kernel functions such as {\tt input}, {\tt output}, or {\tt inout}.}
\REM{StarSs programming model family introduces the ability to extract task parallelism in the presence of data dependences. These models use programmer annotations of input and output operands in tasks (kernel functions) to construct an inter-task data dependence graph dynamically. Calls to tasks are checked at runtime for dependences by analyzing the addresses of their parameters and by using programmer annotations~\cite{EtsionMICRO10}.}

\subsection{OpenMP Tasks}

\REM{Tasks in OpenMP are blocks of code that the compiler envelops and provides to be executed in parallel.}Tasks were added to OpenMP in version 3.0~\cite{AyguadeTPDS09}. In OpenMP 4.0~\cite{openmp2013}, the {\tt depend} clause  and the {\tt taskgroup} construct were incorporated and, in OpenMP 4.5, the {\tt taskloop} construct was proposed~\cite{openmp2015}. \REM{Fig.~\ref{lst:task1} shows the code of a task using OpenMP.}Like work-sharing constructs, tasks must be created inside of a {\tt parallel} region. To spawn each task once, the {\tt single} construct is used. \REM{In the example, a task only prints a message of {\tt hello world} once and another task prints {\tt hello again} once too}The ordering of tasks is not defined, but there are ways to specify it: (a) with directives such as {\tt taskgroup} or {\tt taskwait}; or (b) with task dependences ({\tt depend} clause).

\REM{Variables that are used in tasks can be specified with data-sharing attribute clauses ({\tt private}, {\tt firstprivate}, {\tt shared}, etc.) or, by default, data accessed by a task is {\tt shared}.
The {\tt depend} clause takes a type ({\tt in}, {\tt out}, or {\tt inout}) followed by a variable or a list of variables. These types establish an order between sibling tasks. {\tt taskwait} clause waits for the child tasks of the current task. {\tt taskgroup} is similar to {\tt taskwait} but it waits for all descendant tasks created in the block. Moreover, task {\tt reduction}  was introduced in OpenMP 5.0~\cite{openmp2018}.}


The {\tt taskloop} construct was proposed in~\cite{TeruelIWOMP13} and allows parallelizing a loop by dividing its iterations into a number of created tasks, with each task being assigned to one or more iterations of the loop.\REM{Fig.~\ref{lst:example_taskloop} shows a code which will create 32 tasks in a new implicit {\tt taskgroup}.} The {\tt grainsize} clause specifies how many iterations are assigned  for each task and the number of tasks can be calculated automatically.\REM{ OpenMP  brings another clause called {\tt priority} to specify the level of priority of each task used by the runtime scheduler~\cite{openmp2015}.} {\tt taskloop} is compliant with the {\tt parallel for} construct, the main difference is the lack of the {\tt schedule} clause in the {\tt taskloop}\REM{~\cite{PodobasIWOMP16}}.

%
%

\subsection{TLS on Hardware Transactional Memories}

When a compiler cannot prove that a loop can be executed in parallel but it can estimate with a high probability that the loop iterations will be independent at runtime, it can schedule the parallel execution of the loop speculatively. A mechanism is then necessary to detect when a dependence does occur at runtime (Speculative DOACROSS) and to re-execute the loop iterations that were compromised. This technique is known as Thread-Level Speculation. TLS has been widely studied~\cite{SohiISCA95,SteffanISCA2000}. For performance, TLS requires hardware mechanisms that support four primary features: conflict detection, speculative storage, in-order commit of transactions, and transaction roll-back.  However, to this day, there is no off-the-shelf processor that provides direct support for TLS. Speculative execution is supported, however, in the form of Hardware Transactional Memory (HTM) available in processors such as the Intel Core and the IBM POWER~\cite{SalamancaEURO17}. HTM implements three out of the four key features required by TLS: conflict detection, speculative storage, and transaction roll-back. And thus these architectures have the potential to be used to implement TLS~\cite{SalamancaTPDS18,SalamancaEURO17}. Our implementation is based on this approach.

\subsection{Speculative Privatization}

Speculative privatization is a technique that eliminates some false dependences at the cost of increased memory footprint and runtime checks that validate the safety of data accesses~\cite{JohnsonPLDI12,RauchwergerTPDS99}. It involves costly instrumentation of all memory accesses of privatized objects for logging or communication. At commit, the private copies of each worker are merged according to a resolution policy.

\section{Speculative Task Execution}\label{sec:proposal}

In this section we present the new OpenMP extensions that enable programmers to easily annotate loops that can be then speculatively executed using explicit tasks generated by {\tt taskloop}. These extensions allow programmers to parallelize {\it may} DOACROSS loops and to annotate scalar or array variables to be speculatively privatized.

\subsection{{\tt taskloop} {\tt tls} and {\tt spec\_private} clauses}

The use of the {\tt tls} clause for {\tt taskloop} is syntactically similar to the {\tt grainsize} clause, thus they are mutually exclusive and should not appear on the same {\tt taskloop} directive. The use of the {\tt spec\_private} clause is possible in  {\tt taskloop} constructs when the clause {\tt tls} is present; it is syntactically similar to the standard {\tt private} clause. The syntax is as follows:

\begin{figure}[!h]
\vspace{-7mm}
\begin{lstlisting}[escapeinside={\%*}{*)},basicstyle=\normalsize\ttfamily,numbers=none,frame=none]
#pragma omp taskloop tls(%*{\it strip\_size}*)) spec_private(%*{\it list}*)) %*{\it [clause[[,]clause]...]}*)
   %*{\it for-loop}*)
\end{lstlisting}
\label{lst:taskloop_tls_clause}
\vspace{-5mm}
\end{figure}

\noindent where:

\begin{inparaitem}

\noindent 
\item {\it strip\_size} is the number of iterations assigned to each speculative task generated by {\tt taskloop}. In compiler parlance, it is said that the loop is partitioned into strips, and thus this size is often called  the {\it strip size} of the loop;

\noindent 
\item {\it list} consists of a collection of one or more {\it list items} separated by commas;

\noindent
\item {\it list item} is a scalar variable or an array;

\noindent 
\item {\it clause} can be any clause allowed for {\tt taskloop} except {\tt grainsize}, {\tt num\_tasks}, and {\tt collapse}. 

\end{inparaitem}
\vspace{1mm}

{\tt spec\_private} can be used when the loop has {\it may} loop-carried dependences in {\tt shared} variables that are not privatizable at compile time because of the complexity of the analysis (e.g., pointers) or the indeterminism of the control flow of the program. Static analysis fails when the loop has {\it may} loop-carried dependences that arise only if a certain flow of a program is taken at runtime. For instance, in the loop of Fig.~\ref{lst:loopV_tls}, {\tt n} may need to be  declared as {\tt shared} because, in a conservative way, it is necessary to assume a RAW loop-carried dependence in {\tt n}; however, it could be marked as {\tt spec\_private} (Fig.~\ref{lst:loopV_tls}'s code), thus a copy ({\tt nL}) is created for each task such that it can replace {\tt n} within the transaction. Analogously, in the loop of Fig.~\ref{lst:example_tls_priv}, if the condition {\tt cond} evaluates to false in all iterations, {\tt glob} will be loop private, so the programmer could annotate {\tt glob} to be speculatively privatized using a private copy {\tt globL}.

We also propose the clause {\tt spec\_private} to enable programmers to mark arrays to be speculatively privatized. For example, \REM{Fig~\ref{lst:example_tls_priv} shows the use of the clause {\tt spec\_private}. }{\tt spec\_private(A)} transforms the loop in Fig.~\ref{lst:example_tls_priv}, creating thread-local arrays (e.g., {\tt AL}) to perform writes within the transaction and then non-speculatively copying them back to the original array {\tt A} after committing. 

\begin{figure}[t]
\vspace{-3mm}
%
\begin{lstlisting}[escapeinside={\%*}{*)}]
    #pragma omp parallel num_threads(N_CORES)
    #pragma omp single
    #pragma omp taskloop tls(S_SIZE) firstprivate(n) spec_private(glob,A,B)
    for (i = INI; i < n; i++) {
       #pragma omp tls write(glob)
       if (/*cond*/){ %*\label{line:cond3}*)
          #pragma omp tls if_read(glob)          
          glob++; %*\label{line:glob++3}*)
       }else 
          glob=i; %*\label{line:glob=i3}*)
       #pragma omp tls write(A)
       A[i]= glob*i;   %*\label{line:A[i]3}*)
       if (/*cond2*/)
          #pragma omp tls if_write(B) %*\label{line:if_writeB}*)
          B[i]=glob*glob;         %*\label{line:B[i]3}*)
    }
\end{lstlisting}
\vspace{-3mm}
\caption{Code using {\tt taskloop-tls}, {\tt spec\_private}, and {\tt tls} construct. {\tt cond} and {\tt cond2} depend on the input}
\label{lst:example_tls_priv}
\vspace{-4mm}
\end{figure}

\subsection{{\tt tls} construct}

{\tt tls} construct is a stand-alone directive that specifies if a variable is written or first read in all or some iterations of the loop. Thus, it can be used to specify transient {\it may}-RAW-dependence or false-sharing patterns in loops. 

The use of this directive is possible only when {\tt spec\_private} is present. The syntax of the directive is as follows:

\begin{figure}[!h]
\vspace{-7mm}
\begin{lstlisting}[escapeinside={\%*}{*)},basicstyle=\normalsize\ttfamily,numbers=none,frame=none]
#pragma omp tls %*{\it [clause[[,]clause]...]}*)
\end{lstlisting}
\label{lst:use_tls_directive}
\vspace{-5mm}
\end{figure}

\noindent where:

\begin{inparaitem}

\noindent 
\item {\it clause} is one of the following: (a) {\tt read({\it scalar})}, which specifies that {\it scalar} is read before any write to {\it scalar} for each loop iteration; (b) {\tt write({\it item})}: {\it item} is written for each loop iteration; (c) {\tt if\_read({\it scalar})}: {\it scalar} can be read before any write to {\it scalar} for some loop iterations depending on the {\tt if} control flow; and (d) {\tt if\_write({\it item})}: {\it item} can be write for some loop iterations depending on the {\tt if} control flow.

\noindent
\item {\it item} is a scalar variable or an array;

\noindent
\item {\it scalar} is a scalar variable.

\end{inparaitem}

\vspace{1mm}
For example, in the code of Fig.~\ref{lst:loopV_tls}, the {\tt tls if\_write(n)} directive indicates that the variable {\tt n} can be written in some iterations of the loop depending on the control flow, and it could generate a loop-carried dependence. Thereby, only when the {\tt if} condition is true at runtime, the private copy ({\tt nL}) is non-speculatively copied back to {\tt n} after committing. Moreover, when the {\it may} loop-carried dependence is within conditional statements (e.g., Fig.~\ref{lst:loopV_tls}'s loop or Fig.~\ref{lst:example_tls_priv}'s loop ) and the programmer knows the possible sinks of this dependence (Line~\ref{line:read_n2} in Fig.~\ref{lst:loopV_tls} and Line~\ref{line:glob++3} in Fig.~\ref{lst:example_tls_priv}), he/she can annotate the scalar variable to be speculatively privatized (using {\tt spec\_private}) and to be speculatively read when the conditional statements are true at runtime (using {\tt tls if\_read}).

In the code of Fig.~\ref{lst:example_tls_priv}, the {\tt tls write(glob)} directive indicates that the variable {\tt glob} is actually written in all iterations of the loop and could generate a loop-carried dependence. Thereby, the private copy ({\tt globL}) has to be non-speculatively copied back to {\tt glob} after committing. Analogously, the {\tt write} clause for arrays means that every loop iteration writes array {\tt A}, thereby the non-speculative writes from {\tt AL} to {\tt A} are also performed for every iteration.  On the other hand, {\tt if\_write(B)}, shown in Line~\ref{line:if_writeB} of Fig~\ref{lst:example_tls_priv}, analogously to the case of scalars, indicates that the writes to the array are conditioned to the control flow of the program (in the example, the result of {\tt cond2} in each iteration). Therefore, array {\tt B} is speculatively privatized ({\tt spec\_private(b)}) and, only when the conditional statement is true at runtime, non-speculatively written after committing. 

\begin{figure}[t]
\vspace{-3mm}
\centering
\begin{lstlisting}[escapeinside={\%*}{*)},classoffset=3,morekeywords={BEGIN,END},keywordstyle=\color{black}\underline]
    int next_strip_to_commit=INI %*\label{line:INI}*);
    #pragma omp parallel num_threads(N_CORES)
    #pragma omp single
    #pragma omp taskloop grainsize(1) firstprivate(n) shared(glob,A,B)
    for (i = INI; i < n; i+=S_SIZE) {    
       int globL,flag_r_glob=0,count_1=-1;
       int AL_1_1[S_SIZE],BL_1_2[S_SIZE];
       char pred_B_2[S_SIZE]={0};
       
       int speculative=BEGIN(&next_strip_to_commit,i);
       for (int ii=i; ii < n && ii - i < S_SIZE; ii++){        
         count_1++; %*\label{line:count}*)
         if (/*cond*/){ %*\label{line:cond4}*)
            if (!flag_r_glob){ %*\label{line:begin_if_read}*)
              flag_r_glob=1;
              globL=glob;
            } %*\label{line:end_if_read}*)
            globL++; %*\label{line:glob++4}*)
         }else 
            globL=ii; %*\label{line:glob=i4}*)                    
         
         AL_1_1[count_1]= globL*ii;   %*\label{line:AL=}*)
         if (/*cond2*/){ 
            pred_B_2[count_1]=1;  %*\label{line:begin_BL=}*) 
            BL_1_2[count_1]=globL*globL;            %*\label{line:end_BL=}*)   
         } 
       }      
       END(speculative, &next_strip_to_commit, i);
       glob=globL;  %*\label{line:glob_copy_back}*)
       count_1=-1;       %*\label{line:begin_ns_writes}*)          
       for (int ii=i; ii < n && ii - i < S_SIZE; ii++){        
         count_1++;         
         A[ii]=AL_1_1[count_1];  
         if (pred_B[cont1]);
            B[ii]=BL_1_2[count_1];
       }  %*\label{line:end_ns_writes}*) 
       next_strip_to_commit+=S_SIZE;          
    }    
       
\end{lstlisting}
\vspace{-3mm}
\caption{Code converted to standard OpenMP}
\label{lst:example_openmp}
\vspace{-4mm}
\end{figure}

\subsection{{\tt spec\_reduction} clause}

This clause can be used when the loop has {\it may} loop-carried dependences in {\tt shared} variables that have a pattern of reduction in the loop but the standard OpenMP {\tt reduction} cannot be used because of uncertainty of having cross-iteration dependences at runtime. 

The use of the {\tt spec\_reduction} clause is possible in  {\tt taskloop} constructs when the clause {\tt tls} is present. It is syntactically similar to the standard task {\tt reduction} clause. The syntax of {\it taskloop} is as follows:

\begin{figure}[!h]
\vspace{-7mm}
\begin{lstlisting}[escapeinside={\%*}{*)},basicstyle=\normalsize\ttfamily,numbers=none,frame=none]
#pragma omp taskloop tls(%*{\it strip\_size}*)) spec_reduction(%*{\it reduction-identifier}*):%*{\it list}*)) %*{\it [clause[[,]clause]...]}*)
  %*{\it for-loop}*)
\end{lstlisting}
\label{lst:spec_reduction_clause}
\vspace{-5mm}
\end{figure}

\noindent where:
\begin{inparaitem}

\noindent 
\item {\it reduction-identifier} is one of the following operators: {\tt +}, {\tt -}, {\tt *}, {\tt \&}, {\tt |}, {\tt \^~}, {\tt \&\&} and {\tt ||};

\noindent 
\item {\it list} consists of a collection of one or more {\it scalar} separated by commas;

\noindent 
\item {\it clause} can be any clause allowed for {\tt taskloop} except {\tt grainsize}, {\tt num\_tasks}, {\tt nogroup}, and {\tt collapse}. 
\end{inparaitem}

\begin{algorithm}[!b]
\footnotesize
 \SetArgSty{textup}
 \KwData{{\tt taskloop} construct (directive {\it D} and for-loop {\it L}) and {\it strip\_size}}
 \KwResult{Transformed code to be parallelized with TLS on HTM}
 Create {\tt BEGIN} and {\tt END} functions\;
 Outside of the construct, create a new variable {\it next} whose identifier is {\tt next\_strip\_to\_commit} of the same type of the induction variable\;
 Initialize {\it next} to the initial value of the induction variable\;
 Set {\tt grainsize} to {\tt 1} in {\it D}\;
 Apply strip mining transformation to the loop {\it L} using loop-local variable {\it ii} and a size of strips equal to {\it strip\_size} (induction variable is replaced by {\it ii} in he inner loop {\it L'})\;
 Insert a call to the {\tt BEGIN} function before {\it L'}\;
 Insert a call to the {\tt END} function after {\it L'}\;
 \If{{\it D.list\_spec\_private} $\neq$ NULL}{ 
     {\it flag\_array\_nspec\_w} $\gets$ 0\;
     \ForEach{{\it var} $\in$ {\it D.list\_spec\_private}}
     {
         \lIf{{\it var} is scalar}{Run Spec\_Private\_Scalar\_Algorithm}
         \ElseIf{{\it var} is array}{Run Spec\_Private\_Array\_Algorithm\;
           \lIf{{\it flag\_array\_nspec\_w} $=$ 0} {\\{\it flag\_array\_nspec\_w} $\gets$ 1}
         }
     }
     \lIf{ {\it flag\_array\_nspec\_w}$=$ 1} {Run Algorithm~\ref{alg:tls4}}
 }
 \If{{\it D.list\_spec\_reductions} $\neq$ NULL}{ 
     \lForEach{scalar {\it var} $\in$ {\it D.list\_spec\_reduction}}{Run Algorithm~\ref{alg:tls5}} 
  
 }
 At the end of {\it L}, insert a statement to increment the value of {\it next} by {\it strip\_size}\; 
 \caption{Mechanism for {\tt taskloop tls}}
  \label{alg:tls}
\end{algorithm}

\subsection{Implementation of the clauses}\label{sec:implementation}

Clang 4.0 was adapted to generate the AST ({\it Abstract Syntax Tree}) to support the new clauses. For the following discussion, consider Fig.~\ref{lst:example_openmp}, which shows the OpenMP translated code from Fig.~\ref{lst:example_tls_priv} for didactic reasons. 
The translation mechanism for the clause {\tt tls({\it strip\_size})} is listed in Algorithm~\ref{alg:tls}.

The variable {\tt next\_strip\_to\_commit} is {\tt shared} for the construct and controls the order of transaction commits. This variable is initialized to the value of the first iteration, {\tt INI} in the example of Fig.~\ref{lst:example_openmp} (Line~\ref{line:INI}). Moreover, {\tt grainsize} is set to {\tt 1} and it is the {\tt tls} clause that controls the chunk of iterations that each explicit task will execute. The schedule of tasks depends on the runtime and could be non-monotonic (not in increasing iteration order).

When {\tt spec\_private({\it list})} clause is present in the directive and depending on the type of each variable in the {\it list}, code transformation algorithms to enable speculative privatization of scalar or array can be executed (algorithms not shown due to space reasons). All variables in {\tt spec\_private} are set as {\tt shared}. {\tt Spec\_Private\_Scalar\_Algorithm} creates a private copy for {\tt glob} called {\tt globL} in the example of Fig.~\ref{lst:example_openmp}, which replaces {\tt glob} in the inner loop after applying strip mining. Then, it creates the mechanism specified by the directive {\tt if\_read(glob)} to read {\tt glob} within the transaction only when it is actually read in the {\tt if} statement (Lines~\ref{line:begin_if_read}-\ref{line:end_if_read} in Fig.~\ref{lst:example_openmp}). Because of the variable {\tt glob} is written in every iteration, which is specified by {\tt write(glob)}, {\tt globL} is copied to {\tt glob} after committing (Line~\ref{line:glob_copy_back}). 

\REM{{\tt Spec\_Private\_Array\_Algorithm} goes through a structure that groups array writes by indexes, to these indexes by array variables, and finally these variables by {\tt fors}. For each {\tt for} where the array writes, it creates a counter if this does not exist ({\tt count\_1} in Fig.~\ref{lst:example_openmp}'s example) and a statement to update the counter in each iteration of that {\tt for} (Line~\ref{line:count} in the example).  In each write, the algorithm creates a mechanism to replace the array with the private copy of this. In the case of {\tt if\_write}, it creates  predicates to know if the copy back for a determined position of the array must be performed. In the example of Fig.~\ref{lst:example_openmp}, the algorithm}
{\tt Spec\_Private\_Array\_Algorithm} creates a private copy for {\tt A} of size {\tt S\_SIZE} ({\tt AL\_1\_1}), then it replaces {\tt A} with its copy in the inner loop (Line~\ref{line:AL=}), and it finally copies back the private array to the {\tt shared} array {\tt A} after committing. Analogously, the algorithm proceeds for {\tt B} (Lines~\ref{line:begin_BL=}-\ref{line:end_BL=}), but because it is within an {\tt if} statement, it creates an array of predicates to determine if a position of the array is written or not. Then, these predicates are used to copy back the private array to {\tt B}.

\begin{algorithm}[t]
\footnotesize
 \SetArgSty{textup}
 \KwData{{\tt taskloop} construct and {\it list\_fw}}
 \KwResult{Transformed code with the non-speculative copy back of arrays after committing}
   
   \ForEach{{\it st} $\in$ {\it list\_fw}}{
      \If{{\it st} is for}{
        Create a for statement {\it for\_st} with {\it init}, {\it cond}, and {\it inc} the same of {\it st}\;
        Create a statement to increment the value of {\it st.count} by {\tt 1} and push back it in the body of {\it for\_st}\;
        {\it for\_st.count} $\gets$ {\it st.count}\; 
        Push {\it for\_st} in stack {\it S}\;        
      }
      \ElseIf{{\it st} is end\_for}{
        {\it for\_st2} $\gets$ Pop stack {\it S}\;
        \lIf{ {\it S} $=$ $\O$}{
           At the end of {\it L}, insert {\it for\_st2}
        }
        \lElse{Push back {\it for\_st2} in  the body of {\it S.top}}
      } 
      \ElseIf{{\it st} is write}{  
         Create a statement {\it s2} to assign the value at the position {\it S.top.count.value} of the array {\it st.varL} to array {\it st.var} at position {\it st.index.expr}\;     
         \If{ {\it st.type} $=$ {\tt IF\_WRITE}}{
            Create an if statement {\it if\_st} with condition {\tt (<{\it st.pred.id}>[<{\it S.top.count.id}>])}, and set {\it s2} as the then-part\;
            Push back {\it if\_st} in the body of {\it S.top}\;         
         }\lElseIf{ {\it st.type} $=$ {\tt WRITE}}{ Push back {\it s2} in the body of {\it S.top}}
        
      }
   }
 \caption{Mechanism for non-speculative writes of arrays}
 \label{alg:tls4}
\end{algorithm}

Algorithm~\ref{alg:tls4} uses a statement list ({\tt for} and write statements). which follows the original order in the loop. Basically, the algorithm goes through the statement list in order and uses a stack to simulate the nesting of loops. When it finds a {\tt for}, it creates a new {\tt for} statement with the same initialization, condition and increment as those in the found {\tt for}, then it pushes the created {\tt for} in the stack. Notice that the {\tt for} in which writes are performed will be always at the top of the stack. On the other hand, if it finds an end of {\tt for}, it pops from the stack a {\tt for}; hence, if the stack is empty, it inserts this {\tt for} after the {\tt END} function call. However, if the stack is not empty yet, the popped {\tt for} is inserted at the end of the loop body of the {\tt for} that is at the top of the stack. If the statement is a write, a mechanism is created to copy the private array back to the original {\tt shared} array and,  depending on whether it was declared as {\tt if\_write} or {\tt write}, the presence of predicates may be needed to define which positions of the array are actually copied. Then, these writes are placed at the end of the {\tt for} body  that is at the top of the stack. As shown in Fig.~\ref{lst:example_openmp} (Lines~\ref{line:begin_ns_writes}-\ref{line:end_ns_writes}), private arrays are copied back to {\tt shared} arrays {\tt A} and {\tt B}. In the case of {\tt B}, predicates are used. The algorithm for {\tt spec\_reduction} is shown in Algorithm~\ref{alg:tls5}.

\begin{algorithm}[t]
\footnotesize
 \SetArgSty{textup}
 \KwData{{\tt taskloop} construct, {\it var}, {\it op\_red}, and {\it statement\_red} ({\tt {\it omp\_out} = {\it omp\_out} <{\it op\_red}> {\it omp\_in}} )}
 \KwResult{Transformed code with speculative reduction for {\it var}}
   Set {\it var} as {\tt shared}\;
   Create a variable {\it varL} of the same type of {\it var} (with identifier ``{\tt <{\it var.id}>}" plus ``{\tt L}")\;
   Initialize the value of {\it varL} to the identity value of operator {\it op\_red}\;
   Replace {\it statement\_red.omp\_out} with {\it varL} in the statement {\it statement\_red}\;
   Create a statement {\it st\_copy}, copy of {\it statement\_red}\;
   Replace {\it varL} with {\it var} and {\it st\_copy.omp\_in} with {\it varL}\ in {\it st\_copy}\;
   After the {\tt END} function call, insert {\tt st\_copy}\;
 \caption{Mechanism for {\tt spec\_reduction}}
  \label{alg:tls5} 
\end{algorithm}



\begin{figure}
\vspace{-3mm}
\centering
\includegraphics[scale=0.20]{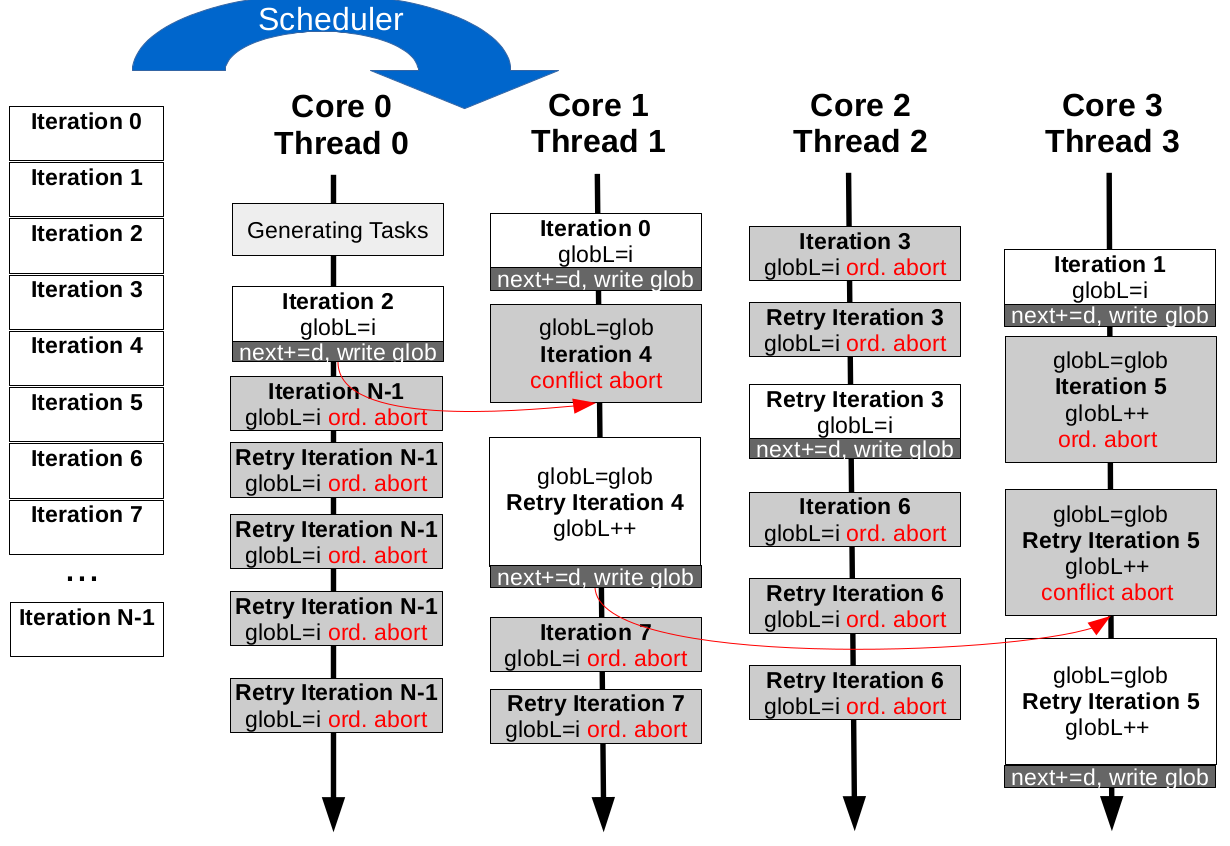}
\vspace{-3mm}
\caption{Possible execution flow of Fig.~\ref{lst:example_tls_priv} with {\tt S\_SIZE}=1 and {\tt N\_CORES}=4}
\label{fig:tls_flow}
\vspace{-4mm}
\end{figure} 

\subsection{How  {\tt taskloop tls} works}

\REM{The {\tt parallel} construct creates a team of OpenMP threads that execute the explicit tasks created by {\tt taskloop}. The number of threads in the example is equal to the number of physical cores because, as explained in previous work~\cite{SalamancaEURO17}, to achieve performance in TLS it is necessary to bound each software thread to one hardware thread since it avoids aborts due to the interference between threads when executing on the same core. With the {\tt single} construct, the {\tt taskloop} construct is executed by only one of the threads in the team. This thread encounters the {\tt taskloop} which partitions the iterations of the loop into explicit tasks which are scheduled at runtime to be executed by the team of threads. 

The {\tt grainsize} clause is set to 1 because if a value greater than 1 is defined the performance degrades since more transactions are created and therefore aborted by the task. As explained in previous work~\cite{SalamancaEURO17}, the overhead of starting, finishing and aborting transactions is high.} Fig.~\ref{fig:tls_flow} shows a possible execution of the {\tt taskloop tls} for the loop of Fig.~\ref{lst:example_tls_priv}. More details about how coarse-grained TLS is implemented on HTM can be found in previous work~\cite{SalamancaTPDS18}.

The {\tt BEGIN} function creates a transaction $T$ that encapsulates {\tt S\_SIZE} speculative iterations. The size of the strip is specified as a parameter of the clause. At runtime, each explicit task created by {\tt taskloop} will execute the {\tt BEGIN} function. On the other hand, the {\tt END} function will attempt to commit the transaction $T$ if all previous strips have already committed and no conflict is detected. Otherwise, the explicit task will abort and re-start $T$. 

\REM{The order of creation of tasks is not specified in OpenMP. The {\tt taskloop} construct does not include a {\tt schedule} clause, although it was proposed by Teruel {\it et al.}~\cite{TeruelIWOMP13}. Therefore, the scheduling of tasks completely depends on the runtime. When using threads, TLS takes advantage of a schedule similar to {\tt static} because it has to be ensure the in-order commits of transactions. When {\tt static} is defined in the {\tt schedule} clause, for {\tt parallel for} for example, the iterations are divided into chunks of a specified size, and these chunks are assigned to threads in a round-robin (in the order of thread numbers) and {\tt monotonic} fashion (in the increasing iteration order). Similarly, depending on the loop and the load balancing of iterations, {\tt dynamic} could work well in TLS but just in a {\tt monotonic} way.}

\begin{table*}[ht]
\vspace{-2mm}
\centering
\caption{Characterization and TLS Execution of Loops.}
\label{tab:results}
\resizebox{\textwidth}{!}{%
\begin{tabular}{c|cccclrcclcc|ccc|c}
\cline{1-5}\cline{7-9}\cline{11-16} 
Loop & 		
\multicolumn{4}{c}{Loop Information} & & 
\multicolumn{3}{c}{Loop Characterization} & & 
\multicolumn{6}{c}{Loop Execution} \\
ID & 
Benchmark&
Location &
\%$Cov$&
Invocations& &
$N$ &  
$\%lc$ & 
Average Iteration &  &
\multicolumn{2}{c|}{taskloop-tls} &
\multicolumn{3}{c}{taskloop-tls + s\_priv} &
\multicolumn{1}{c}{{\tt ordered}} \\ 
&&&&& &  & & Size &  & {\tt S\_SIZE}  & Speed-up  & Spec Priv & {\tt S\_SIZE} & Speed-up & Speed-up  \\ 
\cline{1-5}\cline{7-9}\cline{11-16} 
A  & {\tt automotive\_bitcount} & {\tt bitcnts.c},65 & 100\% & 560 &&  1125000 & 100\% & 12 B &    &  5020  & 0.29  & {\tt spec\_reduction} & 5020  & 1.69   &  0.31 \\
B  & {\tt automotive\_susan\_c} & {\tt susan.c},1458  & 83\% & 344080 && 590 & 0\% & 48 B  &  & 65 & 1.03 & No  & 65  & 1.03  &  0.46 \\

C  & {\tt automotive\_susan\_e} & {\tt susan.c},1118 &  18\% & 165308  && 592 & 0\% & 14 B  &  & 72 & 0.68 & {\tt if\_write} & 72 & 0.87 & 0.37\\
D  & {\tt automotive\_susan\_e} & {\tt susan.c},1057  & 56\% & 166056 && 594 & 0\% & 76 B  &  & 74 & 0.73 & {\tt if\_write} & 74 & 0.85  & 0.38 \\
E  & {\tt automotive\_susan\_s} & {\tt susan.c},725 &   100\% & 22050 && 600 & 0\% & 14 B  &  & 45 & 0.90 & {\tt if\_write} & 25 & 1.87   & 0.87  \\
H  & {\tt automotive\_susan\_e} & {\tt susan.c},1117 &  18\% & 374 && 442 & 0\% & 3 KB &  & 1 & 1.01 & {\tt if\_write} & 1 & 1.86   & 1.57 \\
I  & {\tt automotive\_susan\_e} & {\tt susan.c},1056 &  56\% & 374 && 444 & 0\% & 4 KB &  & 1 & 0.89 & {\tt if\_write} & 1 & 1.26   & 1.03 \\
J  & {\tt automotive\_susan\_s} & {\tt susan.c},723 &  100\% & 49 && 450 & 0\% & 3 KB &  & 1 & 0.23 & {\tt if\_write} & 1 & 0.63  & 0.97    \\ 
V  & {\tt automotive\_susan\_c} & {\tt susan.c},1614 &  7\% & 782 && 440 & 34\% & 1 KB  && 9 & 1.08 & {\tt if\_read}, {\tt if\_write} &  9 & 1.15 & 0.89  \\ 
{\tt mcf}  & {\tt 429.mcf}  & {\tt pbeampp.c},165    & 40\%  & 21854886   &&  300 & 3.1\%  & 300 B  & & 75 & 1.10 & {\tt if\_read}, {\tt if\_write} & 75 & 1.16 & 0.30 \\
\cline{1-5}\cline{7-9}\cline{11-16}  
\end{tabular}
}
\vspace{-1mm}
\end{table*}

\begin{figure*}[t]
\begin{center}$
\begin{array}{ccccc}
\includegraphics[width=1.25in]{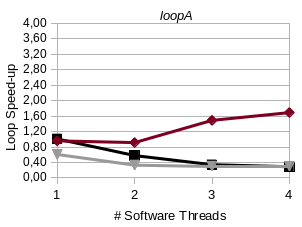} & 
\includegraphics[width=1.25in]{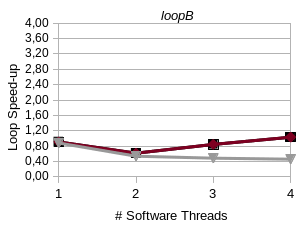} & 
\includegraphics[width=1.25in]{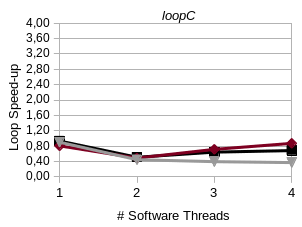} & 
\includegraphics[width=1.25in]{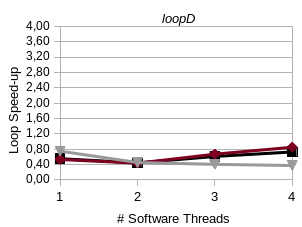} & 
\includegraphics[width=1.25in]{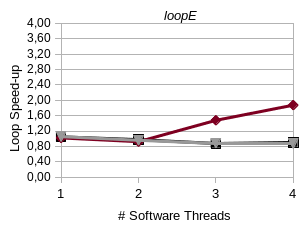} \\
\includegraphics[width=1.25in]{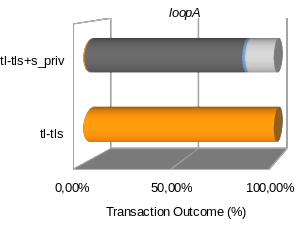}  &
\includegraphics[width=1.25in]{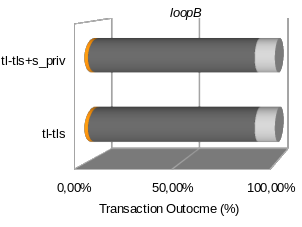}  &
\includegraphics[width=1.25in]{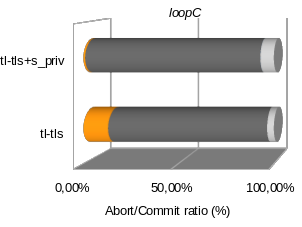}  &
\includegraphics[width=1.25in]{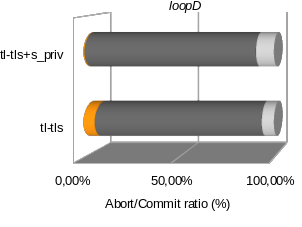}  &
\includegraphics[width=1.25in]{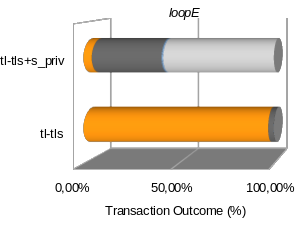}  \\
\includegraphics[width=1.25in]{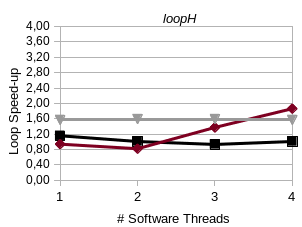} &
\includegraphics[width=1.25in]{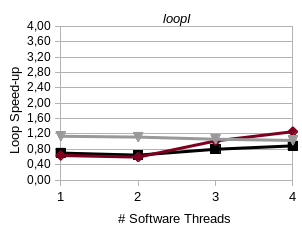} & 
\includegraphics[width=1.25in]{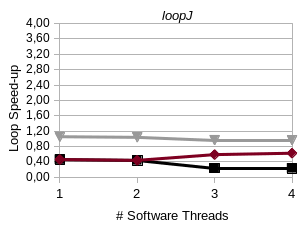} & 
\includegraphics[width=1.25in]{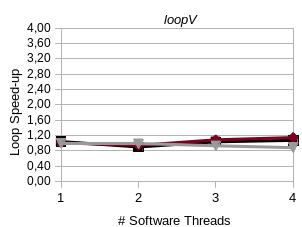} & 
\includegraphics[width=1.25in]{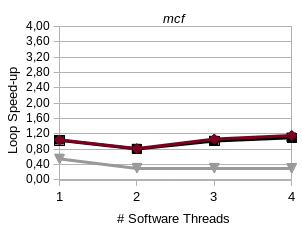} \\
\includegraphics[width=1.25in]{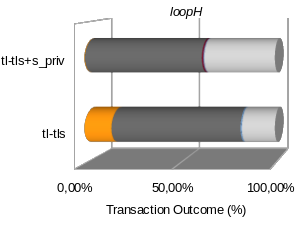}  &
\includegraphics[width=1.25in]{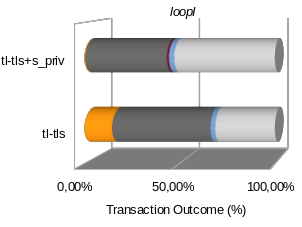}  &
\includegraphics[width=1.25in]{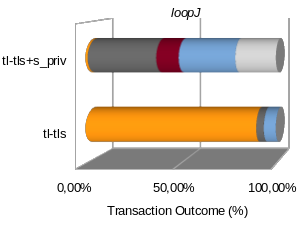}  &
\includegraphics[width=1.25in]{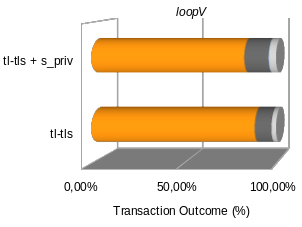}  &
\includegraphics[width=1.25in]{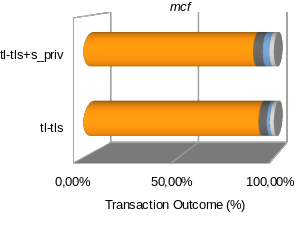} \\
\end{array}$

$\begin{array}{c}
\includegraphics[width=3.3in]{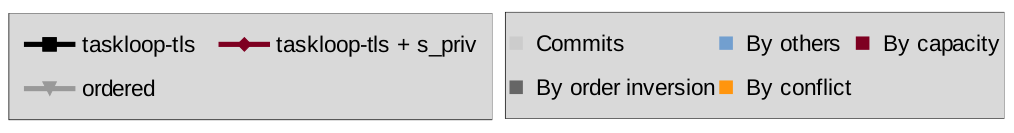} \\
\end{array}
$
\vspace{-2mm}
\caption{Speed-ups and Abort ratios for {\tt taskloop-tls}, {\tt taskloop-tls}+{\tt spec\_priv}, and {\tt ordered} execution on TSX-NI}
\label{fig:speedups}
\end{center}
\vspace{-3mm}
\end{figure*}

The order of creation of tasks is not specified in OpenMP because no kind of schedule is implemented for {\tt taskloop} in OpenMP. This fact could cause a loss of performance in TLS for {\tt taskloop}  since the scheduling could be non-monotonic (Fig.~\ref{fig:tls_flow}), meaning that explicit tasks executing higher iterations could be scheduled before than lower ones. Hence, transactions executed by explicit tasks of higher iterations will be aborted by ordered inversion --- a type of abort where  a transaction that completes execution out of order is rolled-back using an explicit abort instruction ({\tt xabort})~\cite{SalamancaEURO17}. Aborts by order inversion are a problem when using TLS on TSX-NI due to the lack of suspended transactions as previous works showed~\cite{SalamancaEURO17}. However, the problem is exacerbated in task-parallelism model due to non-monotonic scheduling.

\section{Benchmarks, Methodology and Experimental Setup}\label{sec:setup}

The performance assessment in this work reports speed-ups and abort/commit ratios (transaction outcome) for the {\tt taskloop-tls}, {\tt taskloop-tls}+{\tt spec\_private}, and {\tt ordered} parallelizations of {\it may} DOACROSS loops from the Collective Benchmark\cite{cBench} ({\tt cBench}) and SPEC benchmark suites~  running on Intel Core. For all experiments, the default input is used for the {\tt cBench} benchmarks and the reference input for {\tt mcf}. The baseline for speed-up comparisons is the serial execution of the same benchmark program compiled at the same optimization level. Loop times are compared to calculate speed-ups. Each software thread is bounded to one hardware thread (core). Each benchmark was run twenty times and the average time is used. Runtime variations were negligible and are not presented.

Loops were annotated with the {\tt tls} directive as also with the proposed clauses {\tt spec\_private} and {\tt spec\_reduction}, following the syntax described in Section~\ref{sec:proposal}. They were then executed using an Intel Core i7-6700HQ machine, and their speed-ups measured with respect to sequential execution. 
Table~\ref{tab:results} lists the loops used in the study.

This experimental evaluation was carried out on an Intel Core i7-6700HQ processor with 4 cores with 2-way SMT, running at 2.6 GHz, with 16 GB of memory on Ubuntu 18.04.4 LTS (GNU/Linux 4.15.0-112-generic x86\_64). The cache-line prefetcher is enabled by default. Each core has a 32 KB L1 data cache and a 256 KB L2 unified cache. The four cores share an 6144KB L3 cache. 
The benchmarks are compiled with customized Clang 4.0 (it was adapted to generate AST to support the new clauses as explained in Section~\ref{sec:implementation}.) at optimization level -O3 and with the set of flags specified in each benchmark program. Code compiled by {\tt clang -fopenmp} was linked  against the Intel OpenMP Runtime Library. To guarantee that each software thread is bound to one hardware thread (core), the environment variable {\tt KMP\_AFFINITY} is set to {\tt granularity=fine,scatter}.

\section{Experimental Results and Analysis}\label{sec:experiments}

This section presents results and analysis. The first part of Table~\ref{tab:results} shows the information of loops: (1) the ID of the loop in this study; (2) the benchmark of the loop; (3) the file/line of the target loop in the source code; 
(4) $\%Cov$, the fraction of the total execution time spent in the loop; and (5) the number of invocations of the loop in the whole program. 
The features used to characterize the loops are shown in the second part of Table~\ref{tab:results}: (1) $N$, the average number of loop iterations; (2) $\%lc$, the percentage of iterations that have actual RAW loop-carried dependences for the default input of cBench loops and the reference input of {\tt mcf}; and (3) the average size in bytes read/written by an iteration.
The parameters in the third part of Table~\ref{tab:results} describe: (1) {\tt S\_SIZE}, the {\it strip size} used for the experimental evaluation of {\tt taskloop-tls} and {\tt taskloop-tls}+{\tt s\_priv}; (2) the average speed-ups  with four threads for {\tt taskloop-tls}, {\tt taskloop-tls}+{\tt s\_priv}, and  {\tt ordered}; and (3) the type of speculative privatizations (and the clause of {\tt tls} directive) used  in the {\tt taskloop-tls}+{\tt s\_priv} implementation.\REM{of the loops.}

Firstly, it is relevant to note that the performance of {\tt taskloop-tls} would be better if {\tt monotonic} scheduling was set for {\tt taskloop} in the OpenMP runtime. Non-monotonic scheduling causes a task executing higher iterations to be scheduled before others executing lower ones. Thus the former repeatedly abort their transactions due to order inversion. The threads that execute these tasks are rendered unusable for some time (lost-threads effect), bringing down the parallelism that could have been gained using TLS~\cite{SalamancaIWOMP19}. 

Loops {\tt A}, {\tt E}, {\tt H}, {\tt I}, {\tt V}, and {\tt mcf} take advantage of {\tt taskloop-tls} parallelization and the speculative privatization --- it removes almost 100\% of conflict aborts in most loops --- (reduction for {\tt loopA}, scalar for {\tt loopV} and {\tt mcf}, and array for the others),  despite the lost-threads and order-inversion-aborts issues generated by the {\tt taskloop} scheduling. Using two threads, no loop has speed-ups with {\tt taskloop-tls}+{\tt s\_priv}, since as explained in~\cite{SalamancaIWOMP19}, the order-inversion-aborts problem is intensified with decreasing number of threads.

{\tt taskloop-tls}+{\tt s\_priv} performance is poor in loops {\tt B}, {\tt C}, {\tt D}, and {\tt J}. In the first three loops, it is due to the already explained problem of lost threads and order-inversion aborts. In {\tt loopJ}, aborts can be seen due to other causes (OS interrupts), in addition to aborts due to overflow of capacity caused by the speculative privatization carried out in arrays. Besides, {\tt taskloop} can schedule tasks in a non-monotonic way, even only one thread; for example,  loops {\tt D}, {\tt I}, and {\tt J}, as shown in Fig.~\ref{fig:speedups}, have slowdowns using one thread.

Parallelization of some loops (e.g., {\tt A}, {\tt E}, and {\tt J}) using {\tt taskloop-tls} results in quite a few aborts due to conflict caused by false dependences or false sharing. For these loops, as the number of threads decreases, the rate of aborts due to conflict decreases, and better speed-ups are achieved using two threads than with three or four. On the other hand, in the case of {\tt mcf} and {\tt loopV}, the high rate of aborts due to conflict is because {\it may} dependences materialize at runtime. 

{\tt ordered} is suitable when the loop can be statically considered DOACROSS as also its sequential and parallel components can be known --- parallel components have to do significant work concerning the loop iteration time to be performant. If the loop is {\it may} DOACROSS and the presumed sequential component can be known, {\tt ordered} can also be beneficial. However, the most difficult task for the programmer is to recognize these components of the loop.

\begin{figure}[t]
\vspace{-3mm}
\begin{lstlisting}[escapeinside={\%*}{*)}] 
   #pragma omp parallel for ordered(1)... 
   for(i=0; i<iterations; i++){//mcf's loop
      arc=...;
      cond1=arc->ident->BASIC;
      if (cond1)){
         red_cost=...;
         cond2=bea_is_dual_infeasible(arc,red_cost);
      }
      #pragma omp ordered depend(sink:i-1) 
      if (cond1&&cond2){
         basket_size++;
         basket_sizeL=basket_size;
      }
      #pragma omp ordered depend(source)
      if (cond1&&cond2){
         perm[basket_sizeL]->a=...;
         ...
      }
   }
\end{lstlisting}
\vspace{-3mm}
\caption{Restructured {\tt mcf}'s hottest loop to use {\tt depend}}
\label{lst:mcf_ordered}
\vspace{-4mm}
\end{figure}

For example, {\tt mcf}'s loop was the only one of the analyzed loops where it was possible to separate components to make use of fine-grained {\tt ordered}. However, it was needed  a non-trivial loop restructuring (Fig.~\ref{lst:mcf_ordered}) because due to the indeterminism of the control flow and the iterator increment variability,  it is not possible to give a correct value to {\tt depend(sink) basketsize++}. Thus one has to be conservative and assume that the component is sequential for each iteration ({\tt sink:i-1}). In this sense, coarse-grained TLS in tasks (STE) offers the advantage of not needing to recognize the components and speculate the entire iteration, resulting in better performance in most of the loops evaluated.

In general, {\tt ordered}, when applied to evaluated loops, results in poor performance. For the cases of {\tt loopH} and {\tt loopI}, it results in some speed-ups; however, this result is not due to the parallelization itself but to the code transformation by the compiler to implement the {\tt ordered} mechanism.  Proof of this is that the best speed-up with {\tt ordered} is using one thread (without synchronization).

\section{Conclusions}\label{sec:conclusions}

Although DOACROSS Parallelization in OpenMP ({\tt ordered} construct) was only intended for loops whose loop-carried dependences can be known statically in arrays or matrices, there are loops where it is not possible: (a) to identify such dependences; (b) to determine the {\tt sink} iteration because of the indeterminism of the control flow; or (c) to delimit their components by hand due to difficult code readability (difficult pointers or function calls). It is in these loops where STE (TLS + tasks) can exploit the underlying parallelism that the other DOACROSS techniques are not able to identify and have poor performance, achieving effective speculative parallelization and performance improvements. This paper proposes STE through OpenMP clauses for the {\tt taskloop} construct and presents a performance evaluation comparing STE with standard DOACROSS parallelization in OpenMP that reveals that slowdowns using {\tt ordered} can be transformed in speed-ups for some loops.

\bibliographystyle{IEEEtran}
\bibliography{IEEEabrv,references}

\begin{thebibliography}{10}
\providecommand{\url}[1]{#1}
\csname url@samestyle\endcsname
\providecommand{\newblock}{\relax}
\providecommand{\bibinfo}[2]{#2}
\providecommand{\BIBentrySTDinterwordspacing}{\spaceskip=0pt\relax}
\providecommand{\BIBentryALTinterwordstretchfactor}{4}
\providecommand{\BIBentryALTinterwordspacing}{\spaceskip=\fontdimen2\font plus
\BIBentryALTinterwordstretchfactor\fontdimen3\font minus
  \fontdimen4\font\relax}
\providecommand{\BIBforeignlanguage}[2]{{%
\expandafter\ifx\csname l@#1\endcsname\relax
\typeout{** WARNING: IEEEtran.bst: No hyphenation pattern has been}%
\typeout{** loaded for the language `#1'. Using the pattern for}%
\typeout{** the default language instead.}%
\else
\language=\csname l@#1\endcsname
\fi
#2}}
\providecommand{\BIBdecl}{\relax}
\BIBdecl

\bibitem{Lamport1974}
L.~Lamport, ``The parallel execution of do loops,'' \emph{Commun. ACM},
  vol.~17, no.~2, pp. 83--93, Feb. 1974.

\bibitem{CytronICPP86}
R.~Cytron, ``Doacross: Beyond vectorization for multiprocessors,'' in
  \emph{Intern. Conf. on Parallel Processing ({ICPP})}, 1986, pp. 836--844.

\bibitem{HursonAC97}
A.~R. Hurson, J.~T. Lim, K.~M. Kavi, and B.~Lee, ``Parallelization of doall and
  doacross loops—a survey,'' \emph{Advances in computers}, vol.~45, pp.
  53--103, 1997.

\bibitem{VachharajaniPACT07}
N.~Vachharajani, R.~Rangan, E.~Raman, M.~J. Bridges, G.~Ottoni, and D.~I.
  August, ``Speculative decoupled software pipelining,'' in \emph{Parallel
  Architecture and Compilation Techniques ({PACT})}, Brasov, Romania, 2007, pp.
  49--59.

\bibitem{openmp2015}
OpenMP-ARB, ``Openmp application program interface version 4.5,'' 2015.

\bibitem{TeruelIWOMP13}
X.~Teruel, M.~Klemm, K.~Li, X.~Martorell, S.~L. Olivier, and C.~Terboven, ``A
  proposal for task-generating loops in openmp*,'' in \emph{Intern. Workshop on
  OpenMP ({IWOMP})}, Camberra, Australia, 2013, pp. 1--14.

\bibitem{SohiISCA95}
G.~S. Sohi, S.~E. Breach, and T.~N. Vijaykumar, ``Multiscalar processors,'' in
  \emph{International Symposium on Computer Architecture ({ISCA})}, S.
  Margherita Ligure, Italy, 1995, pp. 414--425.

\bibitem{SteffanHPCA98}
J.~Steffan and T.~Mowry, ``The potential for using thread-level data
  speculation to facilitate automatic parallelization,'' in \emph{High-Perform.
  Computer Architecture ({HPCA})}, Washington, USA, 1998, pp. 2--13.

\bibitem{SalamancaIWOMP19}
J.~Salamanca and A.~Baldassin, ``A proposal for supporting speculation in the
  openmp taskloop construct,'' in \emph{Intern. Workshop on OpenMP ({IWOMP})},
  Auckland, New Zealand, 2019, pp. 246--261.

\bibitem{SalamancaTPDS18}
J.~Salamanca, J.~N. Amaral, and G.~Araujo, ``Using
  hardware-transactional-memory support to implement thread-level
  speculation,'' \emph{IEEE Transactions on Parallel and Distributed Systems},
  vol.~29, no.~2, pp. 466--480, Feb 2018.

\bibitem{ShirakoIWOMP13}
J.~Shirako, P.~Unnikrishnan, S.~Chatterjee, K.~Li, and V.~Sarkar, ``Expressing
  doacross loop dependences in openmp,'' in \emph{Intern. Workshop on OpenMP
  ({IWOMP})}, Berlin, Heidelberg, 2013, pp. 30--44.

\bibitem{GayatriHIPC13}
R.~Gayatri, R.~M. Badia, and E.~Aygaude, ``Loop level speculation in a task
  based programming model,'' in \emph{20th Annual International Conference on
  High Performance Computing}, Dec 2013, pp. 39--48.

\bibitem{AyguadeTPDS09}
E.~{Ayguade}, N.~{Copty}, A.~{Duran}, J.~{Hoeflinger}, Y.~{Lin},
  F.~{Massaioli}, X.~{Teruel}, P.~{Unnikrishnan}, and G.~{Zhang}, ``The design
  of openmp tasks,'' \emph{IEEE Transactions on Parallel and Distributed
  Systems ({TPDS})}, vol.~20, no.~3, pp. 404--418, March 2009.

\bibitem{openmp2013}
OpenMP-ARB, ``Openmp application program interface version 4.0,'' 2013.

\bibitem{SteffanISCA2000}
J.~G. Steffan, C.~B. Colohan, A.~Zhai, and T.~C. Mowry, ``A scalable approach
  to thread-level speculation,'' in \emph{Intern. Conf. on Computer
  Architecture ({ISCA})}, Vancouver, British Columbia, Canada, 2000, pp. 1--12.

\bibitem{SalamancaEURO17}
J.~Salamanca, J.~N. Amaral, and G.~Araujo, ``Performance evaluation of
  thread-level speculation in off-the-shelf hardware transactional memories,''
  in \emph{Euro-Par 2017: Parallel Processing}, Santiago de Compostela, Spain,
  2017, pp. 607--621.

\bibitem{JohnsonPLDI12}
N.~P. Johnson, H.~Kim, P.~Prabhu, A.~Zaks, and D.~I. August, ``Speculative
  separation for privatization and reductions,'' in \emph{ACM SIGPLAN Conf. on
  Programming Language Design and Implementation ({PLDI})}, ser. PLDI ’12,
  Beijing, China, 2012, p. 359–370.

\bibitem{RauchwergerTPDS99}
L.~{Rauchwerger} and D.~A. {Padua}, ``The lrpd test: speculative run-time
  parallelization of loops with privatization and reduction parallelization,''
  \emph{IEEE Transactions on Parallel and Distributed Systems ({TPDS})},
  vol.~10, no.~2, pp. 160--180, 1999.

\bibitem{cBench}
cTuning Foundation, ``cbench: Collective benchmarks,
  http://ctuning.org/cbench,'' 2016.

\end{thebibliography}

\end{document}